\documentclass[preprint,superscriptaddress,preprintnumbers]{revtex4}

\usepackage{eurosym}
\usepackage{graphicx}
\usepackage{dcolumn}
\usepackage{bm}
\usepackage{xcolor}
\usepackage{multirow}

\begin{document}

\title{Charge Fluctuations in the NdO$_{1-x}$F$_{x}$BiS$_{2}$ Superconductors}
\author{Anushika Athauda}
\affiliation{Department of Physics, University of Virginia, Charlottesville, VA 22904, U.S.A.}
\author{Yoshikazu Mizuguchi}
\affiliation{Department of Physics, Tokyo Metropolitan University, Tokyo, 192-0397, Japan}
\author{Masanori Nagao}
\affiliation{University of Yamanashi, Kofu 400-8511, Japan}
\author{Joerg Neuefeind}
\affiliation{Oak Ridge National Laboratory, Oak Ridge, TN 37831, U.S.A.}
\author{Despina Louca}
\thanks{Author to whom correspondence should be addressed: louca@virginia.edu%
}
\affiliation{Department of Physics, University of Virginia, Charlottesville, VA 22904, U.S.A.}
\date{\today}

\begin{abstract}
The local atomic structure of superconducting NdO$_{1-x}$F$_{x}$BiS$_{2}$ (\textit{x}=0.2 and 0.4) is investigated using neutron diffraction and the pair density function analysis technique. In the non-superconducting $x$=0.2 composition, ferrodistortive displacements of the pyramidal sulfur ions break the tetragonal symmetry and a superlattice structure emerges with peaks appearing at $h+k$ odd reflections superimposed on the even reflections of the $P4/nmm$ symmetry. In the superconducting $x$=0.4 composition, similar ferrodistortive displacements are observed but with different magnitudes coupled with in-plane Bi distortions. The distortions are indicative of charge fluctuations that may lead to charge localization and suppression of superconductivity.   
\end{abstract}

\maketitle




\section{Introduction}

The new phonon mediated BiS$_2$ superconductors, LnO$_{1-x}$F$_{x}$BiS$_{2}$ (Ln=La, Ce, Nd, Pr and Yb)\cite{Mizu2012Nov,Yazici2013} and Bi$_6$O$_4$S$_4$(SO$_4$)$_{1-x}$\cite{Mizu2012}, have attracted considerable attention recently because of the interplay of spatial disorder, phonon anomalies and superconductivity. Charge carriers are introduced either by electron doping or by defects in the spacer layers that sandwich the BiS$_{2}$ superconducting planes.\cite{Mizu2012,Jha2014} Their crystal structure and possibly strong Fermi surface nesting render them quite similar to the high T$_c$ iron based superconductors.\cite{Mazin2008,Dong2008}  They were theoretically predicted to have a one dimensional charge density wave (CDW) instability that is strongly linked to the structural properties and gives rise to a large electron-phonon coupling. First principles calculations indicated that in LaOBiS$_{2}$, unstable soft phonons occur at the $\Gamma$ point while with 50$\%$ doping at the oxygen site, phonon softening occurs along the entire Brillouin zone, (q,q,0), with unstable phonon modes appearing at the M point as a result of an in-plane sinusoidal distortion of the S and Bi atoms that make up the superconducting planes. The close proximity of superconductivity to a CDW instability may bring about peculiar states that may be chemically tuned and would in turn allow us to probe the nature of the soft phonons and the ground state properties. \cite{Yildirim2013,Wan2013} Moreover, electronic correlations resulting from structural inhomogeneities may promote novel phases that are vital to understand the evolution from the normal to the superconducting state. It is the case in the BiS$_2$ superconductors that disorder can tune and induce new correlated electron phenomena and can lead to electron localization. In this study, the local microstructure of NdO$_{1-x}$F$_{x}$BiS$_{2}$ is investigated to understand the nature of local atomic correlations as a function of $x$ and of their role in electron-lattice interactions.

It is understood by now that LnO$_{1-x}$F$_{x}$BiS$_{2}$ (Ln = La, Nd, Pr, Ce and Yb) shows enhanced superconductivity either under high pressure synthesis during which crystallinity is suppressed or by introducing chemical pressure through substitutions at the Ln site.\cite{Mizu2015,Kaji2014} NdO$_{1-x}$F$_{x}$BiS$_{2}$, the system of interest in this study exhibits a maximum T$_c$ of $\sim$5.5 K at \textit{x}$\sim$0.4. It is the highest transition temperature of a BiS$_{2}$ superconductor under ambient pressure synthesis conditions.\cite{Demura2013,Mizu2015} The layered crystal structure is shown in Fig. 1(a).  Superconductivity is induced by fluorine (F) doping at the oxygen (O) site in the NdO spacer layers. The superconducting sheets consist of BiS$_2$ square pyramids with corner (S1) and apical (S2) sulfur atoms. The crystal structure is nominally tetragonal with the \textit{P4/nmm} space group. Superconductivity is easily destroyed when a magnetic field, H, is applied parallel to the \textit{c}-axis while superconductivity survives when H is applied parallel to the \textit{ab}-plane which gives rise to a large magnetic anisotropy $\gamma$=H$^{ab}$$_{c2}$/H$^c$$_{c2}$$\sim$30-40, comparable to that of another class of Bi-based superconductors, the Bi-based cuprates. \cite{Mizu2015,Nagao2013,Liu2014} H$_{c2}$ is the upper critical field.  The superconducting gap, $\Delta$$_{s}$, to T$_c$  ratio yields a value of 2$\Delta$$_{s}$/k$_{B}$T$_c$ =16.8 \cite{Liu2014} which is much larger than the theoretical value of 3.5 in the weak coupling limit of the Bardeen-Cooper-Schrieffer (BCS) model of type I superconductivity. 

The nature of the crystal disorder is a key component to understand superconductivity in this system.\cite{Fisher1991,John1975,Seo2014,Kle1988} In this work, we investigated the characteristics of the local lattice distortions in NdO$_{1-x}$F$_{x}$BiS$_{2}$ at two compositions, one non-superconducting, $x$=0.2, and one superconducting, $x$=0.4 (Fig. 1(b)), using neutron powder diffraction, and the pair density function (PDF) analysis, combined with X-ray single crystal diffraction. It is observed that in the non-superconducting composition, in-plane displacements of the S1 atoms lead to ferrodistortive lattice modes, while in the superconducting composition, Bi displacements are additionally observed that lead to more complex local modes. At the same time, the apical S2 is displaced along the \textit{c}-axis toward the Nd/O layer in the superconducting sample alone, very similar to what we previously observed in the superconducting LaO$_{0.5}$F$_{0.5}$BiS$_{2}$. The local model used to explain the distortions in superconducting NdO$_{0.6}$F$_{0.4}$BiS$_{2}$ can also explain the formation of a superlattice pattern observed from single crystal X-ray diffraction data of superconducting NdO$_{0.7}$F$_{0.3}$BiS$_{2}$. A possible interplay of the distortions and carrier localization in the BiS$_2$ superconducting planes is discussed. 

\section{Methods}

Polycrystalline samples of NdO$_{1-x}$F$_{x}$BiS$_{2}$ ($x$ =0.2 and 0.4) were prepared using solid-state reaction using Bi grains, Bi$_{2}$S$_{3}$, Nd$_{2}$O$_{3}$, BiF$_{3}$ and Nd$_2$S$_{3}$ powders. Stoichiometric amounts of each composition were ground, palletized and heated in an evacuated quartz tube at 700$^\circ$C for 15 hours. \cite{Omachi2014} Bulk superconductivity with a T$_{c}$=5 K is observed for \textit{x}=0.4 as shown in Fig. 1(b).  The single crystal samples of NdO$_{0.7}$F$_{0.3}$BiS$_{2}$ were prepared using high temperature CsCl/KCl flux method in a vacuum sealed quartz tube with the starting materials of Nd$_2$S$_3$, Bi, Bi$_2$S$_3$, Bi$_2$O$_3$, BiF$_3$, CsCl, and KCl at approximately 616$^\circ$C. \cite{Nagao2013} The powder diffraction measurements were carried out at the Spallation Neutron Source of Oak Ridge National Laboratory using the Nanoscale Ordered Materials Diffractometer (NOMAD). The structure function, S(Q), obtained from the diffraction data was Fourier transformed to obtain the PDF, $\rho$(r). A momentum transfer of Q=35 \AA$^{-1}$ was used for the Fourier transform. The PDF provides a real-space representation of atomic pair correlations and is sensitive to atomic deviations from the average symmetry. The PDF peaks correspond to the probability of finding a pair of atoms seperated by a distance \textit{r}, by averaging snapshots of pairs over time. The same data were also analyzed in reciprocal space using the Rietveld method and the average structure can be well described using the known \textit{P4/nmm} symmetry. The synchrotron X-ray experiment on NdO$_{0.7}$F$_{0.3}$BiS$_{2}$ single crystal was performed at 11-ID-C using X-rays of 105 KeV at the Advanced Photon Source at Argonne National Laboratory. The data were collected at a wavelength of 0.11798 \AA\ in transmission mode.

\begin{figure}[th]
	\includegraphics[width=0.9\textwidth]{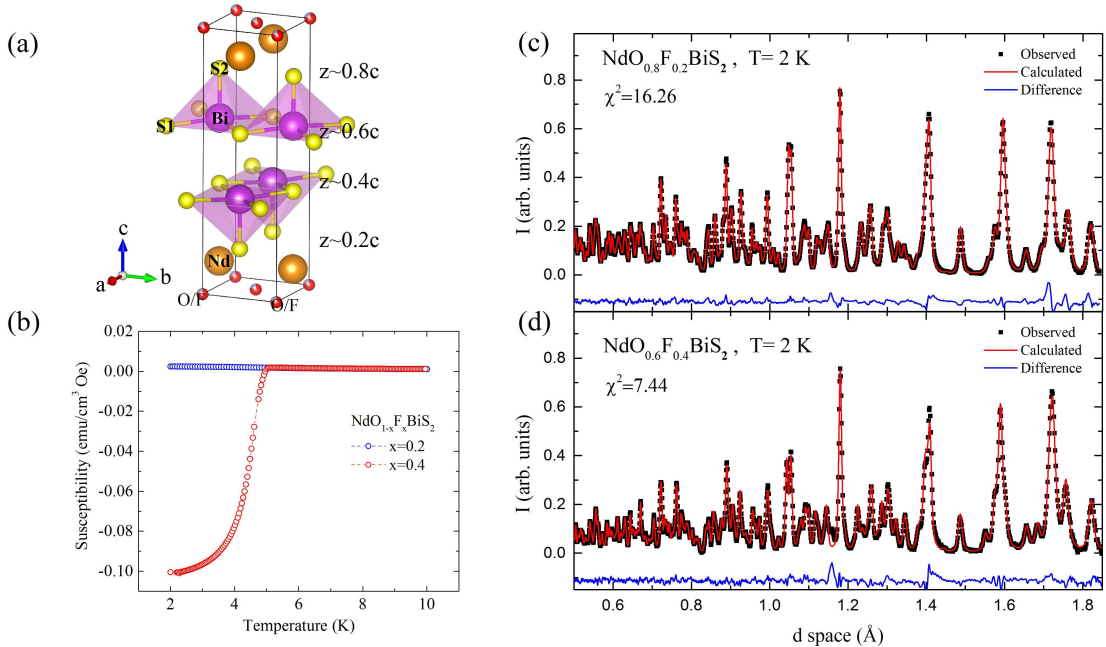} \centering
	\caption{(Color online) In (a), the
		layered crystal structure of NdO$_{1-x}$F$_{x}$BiS$_{2}$ is shown. In (b), the zero-field cooled (ZFC) magnetic susceptibility data of NdO$_{1-x}$F$_{x}$BiS$_{2}$ at non-superconducting \textit{x}=0.2 and superconducting \textit{x}=0.4 are shown. The neutron powder diffraction patterns are shown for NdO$_{1-x}$F$_{x}$BiS$_{2}$ in (c) at \textit{x}=0.2 and (d) at \textit{x}=0.4.}
	\label{FIG.1}
\end{figure}

\begin{figure}[ht]
	\includegraphics[width=0.85\textwidth]{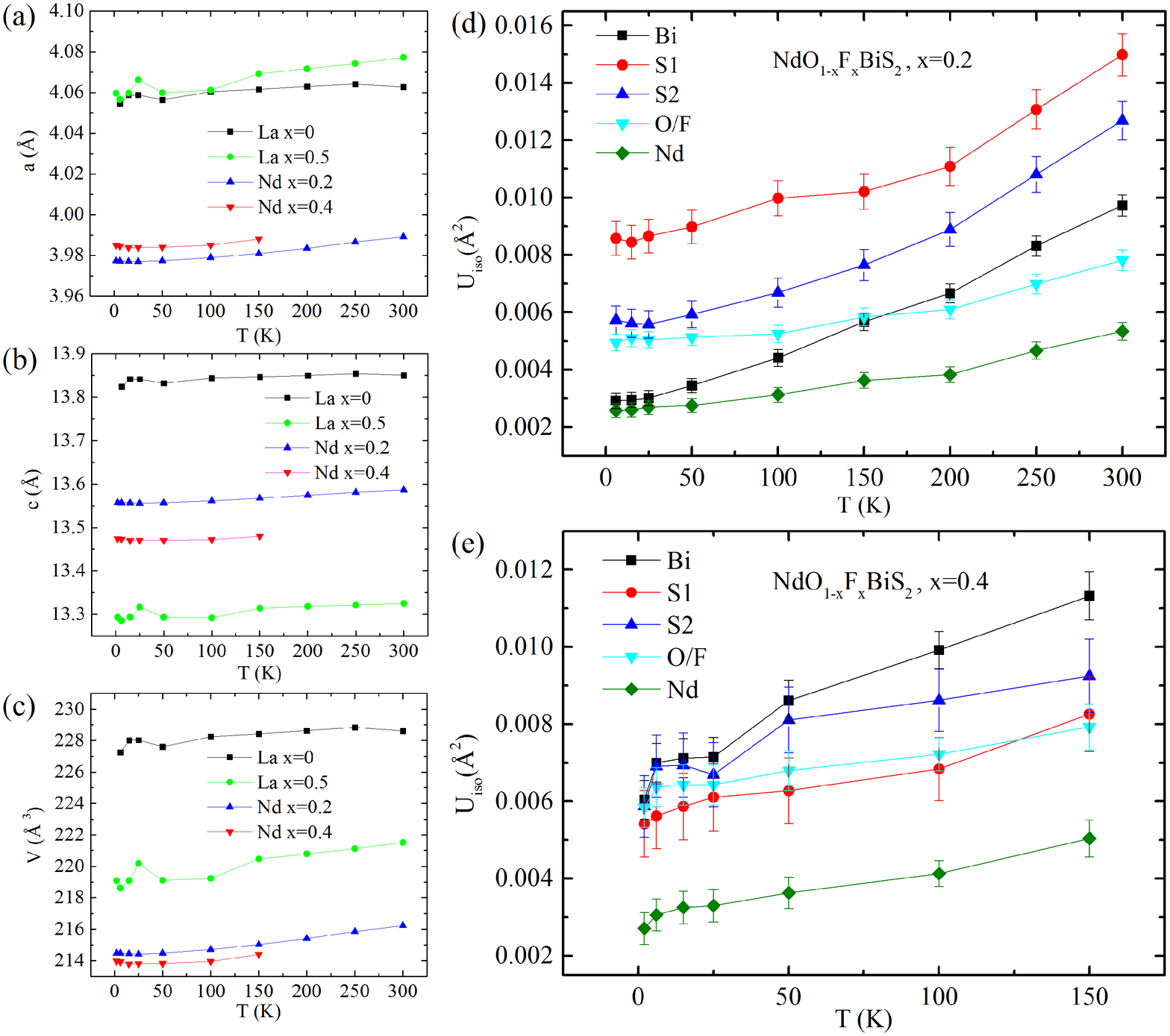} \centering
	\caption{(Color online)  The change of lattice constant \textit{a}, \textit{c} and the unit cell volume as a function of temeperature is shown in (a),(b) and (c) respectively for ReO$_{1-x}$F$_{x}$BiS$_{2}$ (Re=La and Nd). The isotropic thermal factors of NdO$_{1-x}$F$_{x}$BiS$_{2}$ is shown for (d) \textit{x}=0.2 and (e) \textit{x}=0.4.}
	\label{FIG.2}
\end{figure}

\section{Results and Discussion}

The results from the Rietveld refinement of the diffraction data are provided in Table I and also plotted in Fig. 2. The fitting of the diffraction pattern with the model calculated using $P4/nmm$ symmetry is shown in Figs. 1 (c) and (d) for \textit{x}=0.2 and 0.4 at 2 K, respectively. The NOMAD is a medium-range resolution instrument ($\frac{\Delta d}{d}\sim$0.8$\%$) \cite{Jorg2012} and within this resolution, no Bragg peak broadening is observed unlike the case of high pressure annealed LaO$_{1-x}$F$_x$BiS$_2$ in which $(00l)$ peaks were severely compromised. \cite{Lee2013,Athauda2015} The diffraction pattern can be fit well with the \textit{P4/nmm} symmetry for both superconducting and non-superconducting compositions. As a function of $x$, the \textit{c}/\textit{a} ratio decreases from \textit{c}/\textit{a}=3.4087 for \textit{x}=0.2 and \textit{c}/\textit{a}=3.3813 for \textit{x}=0.4 which is due to the $a$ lattice constant expansion as shown in Fig. 2(a) and to the $c$ lattice constant contraction as shown in Fig. 2(b). The $a$- and $c$-lattice constants are compared to those from LaO$_{1-x}$F$_x$BiS$_2$ at $x$=0 and 0.5. Also shown in Fig. 2(c) is a composition of the unit cell volume for the Nd and La compounds. A much larger change is observed in the La doped samples than in the Nd ones. It is clear that the lattice contracts with the replacement of Nd for La, giving rise to a negative chemical pressure. Shown in Figs. 2 (d) and (e) are plots of the isotropic thermal factors, U$_{iso}$, as a function of temperature for \textit{x}=0.2 and 0.4. Similar trends are observed in both except for the Bi-U$_{iso}$ which is larger in the superconducting sample and the U$_{iso}$-for S1 which is larger in the non-superconducting sample. 

\begin{table}[]
	\centering
	\caption{A list of the crystal structure parameters of NdO$_{1-x}$F$_{x}$BiS$_{2}$ at 2 K. O/F is at 2c(0,0,0).}
	\label{Table 1}
	\begin{tabular}{cccccccc}
		\hline
		\multirow{2}{*}{}  & \multirow{2}{*}{R$_{wp}$}    & \multirow{2}{*}{$a$(\AA)}    & \multirow{2}{*}{$c$(\AA)}           & \multirow{2}{*}{\begin{tabular}[c]{@{}c@{}}Nd\\ 2c (0.5,0,z)\end{tabular}} & \multirow{2}{*}{\begin{tabular}[c]{@{}c@{}}Bi\\ 2c (0,0.5,z)\end{tabular}} & \multirow{2}{*}{\begin{tabular}[c]{@{}c@{}}S1\\ 2c (0.5,0,z)\end{tabular}} & \multirow{2}{*}{\begin{tabular}[c]{@{}c@{}}S2\\ 2c (0.5,0,z)\end{tabular}} \\
		&                         &                            &                              &                                                                            &                                                                            &                                                                            &                                                                            \\ \hline
		\multirow{2}{*}{$x$=0.2} & \multirow{2}{*}{0.0355} & \multirow{2}{*}{3.9773(2)} & \multirow{2}{*}{13.5574(8)}  & \multirow{2}{*}{0.09183(7)}                                                & \multirow{2}{*}{0.37178(7)}                                                & \multirow{2}{*}{0.3837(2)}                                                 & \multirow{2}{*}{0.8142(2)}                                                 \\
		&                         &                            &                              &                                                                            &                                                                            &                                                                            &                                                                            \\ \hline
		\multirow{2}{*}{$x$=0.4} & \multirow{2}{*}{0.0576} & \multirow{2}{*}{3.9849(3)} & \multirow{2}{*}{13.4741(11)} & \multirow{2}{*}{0.0941(1)}                                                 & \multirow{2}{*}{0.3735(1)}                                                 & \multirow{2}{*}{0.3825(2)}                                                 & \multirow{2}{*}{0.8145(3)}                                                 \\
		&                         &                            &                              &                                                                            &                                                                            &                                                                            &                                                                            \\ \hline
	\end{tabular}
\end{table}

\begin{figure}[ht]
	\includegraphics[width=0.75\textwidth]{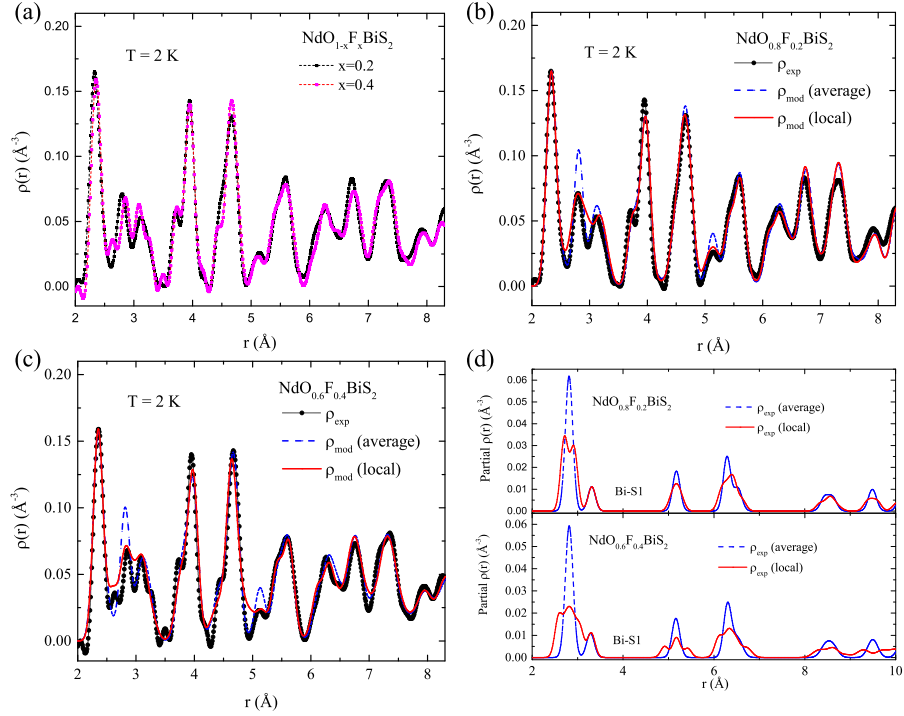} \centering
	\caption{(Color online) (a) A comparison of the data at \textit{x}=0.2 and \textit{x}=0.4 at 2 K. The comparison of average model (in blue dashed line) and local model (in red solid line) with data represented by black dots in (b) NdO$_{0.8}$F$_{0.2}$BiS$_{2}$ and (c) NdO$_{0.6}$F$_{0.4}$BiS$_{2}$. The partial pair density function of Bi-S1 is shown in (d) for \textit{x}=0.2 (top) and \textit{x}=0.4 (bottom).}
	\label{FIG.3}
\end{figure}

The PDFs corresponding to the local structures of \textit{x}=0.2 and \textit{x}=0.4 are plotted in Fig. 3(a). While their local structures are quite similar overall, differences can be seen between the two, especially in the 2.5-3.5 \AA\ range that cannot be accounted for by the substitution of F for O. The \textit{x}=0.2 data is compared to a model PDF calculated using the atomic coordinates and unit cell dimensions obtained from the Rietveld refinement corresponding to the average structure. As shown in Fig. 3(b), overall, the average model $\rho_{mod}$(r) for the $x$=0.2 in the blue dashed line is in good agreement with the experimental PDF, $\rho_{exp}$, except in the 2.5-3.5 \AA\ region where the model peak intensity is higher. The agreement factor, \textit{A}, calculated using $A^2=(n\rho_0^2)^{-1}\sum\limits_{i=1}^n[\rho_{exp}(r_i)-\rho_{mod}(r_i)]^2$  \cite{Toby1992} is 0.181 where $\rho_0$ is the atomic density of the unit cell. On the other hand, a local model, $\rho_{mod}$, involving in-plane displacements at both sulfur S1 and S2 atoms, reproduces the data well as seen in the figure. The A-factor in this case is reduced to 0.162 and the $\rho_{mod}$ is shown as the red solid line in Fig. 3(b). Significant improvement is observed in the 2.5-3.5 \AA\ range. To reproduce the data, the S1 and S2 are displaced by $\sim$0.15 \AA\ and $\sim$0.11 \AA\, respectively, in either the x or y directions away from their equilibrium positions as shown in Fig. 4(a) and (b). There are four planes containing sulfur: S2 at $\sim$0.2c and 0.8c and S1 at $\sim$0.4c and 0.6c as shown in Fig. 1 (a). The in-plane S1 motion gives rise to two different Bi-S1 bond lengths. One possible local model is shown in Fig. 4(a). The S1 atoms at z$\sim$0.6c are displaced in -x direction and at z$\sim$0.4c, they are displaced in the -y direction. The in-plane (-x,-y) mode is not uniquely determined from the powder data because the model only takes into account the magnitude of the bond but not the direction. Thus, equivalent models with (+x,-y)(+x,+y) and (-x,+y) can all produce similar results. The displacements break the four-fold, two fold and inversion symmetries of the \textit{P4/nmm} symmetry. The partial PDFs corresponding to Bi-S1 pair correlation are plotted in Fig. 3(d), and demonstrate how the biggest changes observed in the 2.5-3.5 \AA\ range arise from this pair.

\begin{figure}[ht]
	\includegraphics[width=1.0\textwidth]{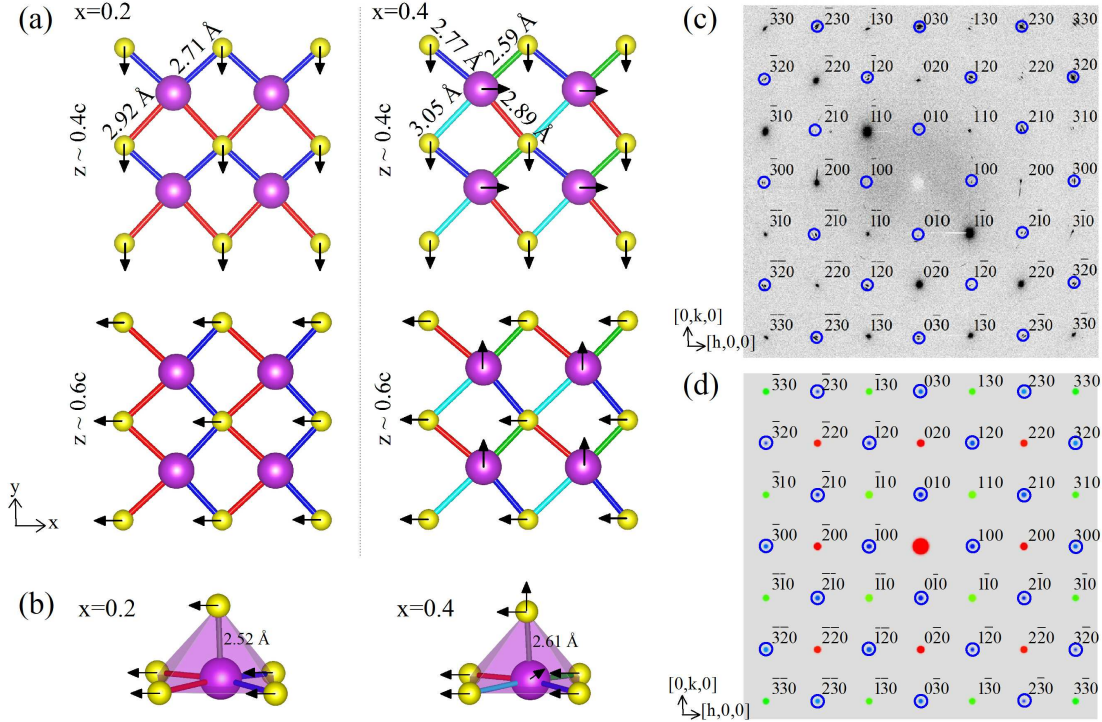} \centering
	\caption{(Color online) (a) The schematics of BiS$_{2}$ planes at 2 K depicting the S1 and Bi motion in NdO$_{1-x}$F$_{x}$BiS$_{2}$, creating two and four magnitudes of bond length at \textit{x}=0.2 and \textit{x}=0.4 respectively. (b) Depicts the in-plane and out of plane displacements of S and Bi at \textit{x}=0.2 and \textit{x}=0.4. (c) The single crystal x-ray diffraction pattern in the hk0 reciprocal plane for NdO$_{0.7}$F$_{0.3}$BiS$_{2}$. (d) The constructed hk0 plane using the local model parameters of NdO$_{0.6}$F$_{0.4}$BiS$_{2}$.} \label{FIG.4}
\end{figure}

The \textit{x}=0.4 data is also compared to a model PDF corresponding to the average structure for the $x$=0.4 composition at 2 K and shown in the blue dashed line in Fig. 3(c). Just as in \textit{x}=0.2, the average model cannot adequately describe the intensity of the peaks in the 2.5-3.5 \AA\ range. To fit the data, a local model $\rho$$_{mod}$ that takes into account in-plane atomic displacements of S1, S2 and Bi and \textit{c}-axis displacement of S2 is used to fit the data and can reproduce the local atomic structure well, as shown by the red solid line in Fig. 3(c). The magnitude of in-plane displacements of Bi, S1 and S2 are $\sim$ 0.12, 0.21 and 0.13 \AA\, respectively. The magnitude of the in-plane displacements of S1 is the same at both planes and so is the magnitude of displacements of S2. The S2 atom is also displaced along c by 0.07 \AA\ as shown in the schematic of the Bi-S tetrahedron in Fig. 4(b). The S1 displacement in $x$=0.4 gives rise to two different Bi-S1 bond lengths similarly to the $x$=0.2, which further split with the addition of the Bi displacement in the $x$=0.4 as shown in Fig. 4(a) and (b). The displacement of Bi gives rise to four different bond lengths in $x$=0.4 as shown in the Bi-S1 partials calculated from the local model and shown in Fig. 3(d). The Bi motion is perpendicular to the S1 motion. It is not peculiar to observe such an asymmetric bond configuration around Bi. Similar observations of asymmetric configurations around Bi have been observed in other systems such as Bi$_{2}$S$_{3}$ and Cu$_{4}$Bi$_{5}$S$_{10}$. \cite{Lun2005, Olsen2010} In contrast, in the non-superconductor, the Bi atom is not displaced from its equilibrium position. The four different Bi-S1 bond lengths do not fluctuate with decreasing temperature.

Indication that the \textit{P4/nmm} symmetry is broken is also provided from the single crystal diffraction data from an \textit{x}=0.3 crystal as shown in Fig. 4(c). The $hk0$ diffraction plane shows the systematic presence of additional Bragg peaks that are not reproduced by the \textit{P4/nmm} symmetry (circled). Using the local model parameters, the $(hk0)$ diffraction pattern is calculated as shown in Fig. 4(d). With the proposed distortion pattern, all reflections not accounted for by the \textit{P4/nmm} symmetry can be reproduced by the local model.

\section{Conclusions}

The crystal structure of NdO$_{1-x}$F$_{x}$BiS$_{2}$ is locally distorted. The distortions are centered around the Bi ion in the superconducting planes and they are very similar in nature to the distortions observed in LaO$_{1-x}$F$_{x}$BiS$_{2}$ with the exception that Bi is displaced from its equilibrium position. \cite{Athauda2015April, Athauda2017} This finding is unexpected because the LaO$_{1-x}$F$_{x}$BiS$_{2}$ system was prepared under high pressure while NdO$_{1-x}$F$_{x}$BiS$_{2}$ is not. The supperlattice Bragg peak formation observed in single crystals of NdO$_{1-x}$F$_{x}$BiS$_{2}$ can be explained by the local model that describes the powder data.

The charge fluctuations around Bi observed indirectly by their effects on the lattice may have important implications on superconductivity. They may be the driving force for the sulfur displacements, the latter of which self organize into (x, y) modes, breaking rotational and the inversion symmetries. The questions to ask are does the superconducting state emerges from the collapse of a long-range CDW state and to what extent are the distortions relevant to pairing? It seems that even in the parent compound, the normal ground state consists of an underlying supperlattice structure that remains even when the system becomes superconducting.

\section{Acknowledgement}

The authors would like to thank Chunruo Duan for help with the data collection and reduction process. This work has been supported by the National Science Foundation, Grant No. DMR-1404994. Work at ORNL was supported by the U.S. Department of Energy, Office of Basic Energy Sciences, Materials Sciences and Engineering Division and Scientific User Facilities Division.

\end{document}